\documentstyle[11pt,aaspp4]{article}
\newcommand{\etal}{{\it et al.\ }}

\begin{document}

\title{A Statistical Examination of the Short Term Stability of the Upsilon Andromedae Planetary System}
\author{Rory Barnes and Thomas Quinn}
\affil{Dept. of Astronomy, University of Washington, Box 351580, Seattle, WA, 98195-1580}
\begin{center}E-mail:rory@astro.washington.edu
\end{center}

\abstract Because of the high eccentricities ($\sim0.3$) of two of the
possible planets about the star Upsilon Andromeda, the stability of
the system requires careful study.  We present results of 1000 numerical
simulations which explore the orbital parameter space as constrained
by the observations.
The orbital parameters of each planet are
chosen from a Gaussian error distribution, and the resulting
configuration is integrated for 1,000,000 years.  We find that 84\% of
these integrations are stable. Configurations in which the
eccentricity of the third planet is $\lesssim0.3$ are always stable,
but when the eccentricity is $\gtrsim0.45$ the system is always
unstable, typically producing a close encounter between the second and
third planets.  A similar exercise with the gas giants in our own Solar
System sampled with the same error distribution was performed. Approximately 81\% of these simulations were stable for $10^6$ years.

\section{Introduction}
The recent observation of three extra-solar planets about the F8 star
Upsilon Andromedae provides a new opportunity to study planetary system
stability. The system consists of the primary and three planets, b, c,
and d, adopting the nomenclature of Butler \etal 1999
(hereafter BMF). Planet b was discovered in 1997 (Butler \etal 1997). The report of two more companions was announced in BMF. This discovery has since been confirmed independently (Noyes \etal 1999).

The implications of this discovery are obvious. New planetary systems
provide opportunities to explore planet formation and nonlinear
dynamics, and increase the probability for both the existence
and detection of life. Until the discovery of the $\upsilon$ And
system, numerical integrations of planetary systems around other stars was
strictly hypothetical (i.e.\ Chambers \etal 1996). Planet formation scenarios must explain hot
Jupiters and highly eccentric planets. With the explosion in the
number of known planets, these fields will experience a revolution in
the coming years.

Much work has already been completed on this system, notably a gigayear
integration (Laughlin and Adams 1999, hereafter LA99), an examination of
possible planets in the habitable zone (Rivera and Lissauer 1999),
integrations of a small sampling of parameter space (Noyes \etal
1999), and simulations by Rivera and Lissauer which explore numerous possibilities in the $\upsilon$ And system (Rivera and Lissauer 2000, hereafter RL00). LA99 integrate only the outer 2 planets of the $\upsilon$
And system. Because the inner planet is the least massive and
is extremely close to the primary, to first order its effects can be
ignored. Removing it changes the dynamical timescale of the system by
two orders of magnitude, making long-term integration feasible. LA99
compensate for the mass/inclination degeneracy by starting the system with a small relative
inclination and set $M_x$ = $M_x\sin i$ ($x$ = b,c,d). This should not
affect the simulation outcome as the transient terms will die out and
the inclinations will approach their natural values. RL00 ran 7
simulations varying timesteps (1/20 to 1/80 of planet b's period),
method of integration, and mutual inclinations. As with LA99, they
find that some configurations eject a planet within $10^5$ years,
while others are stable for $10^8$ years. RL00 also place $\sim300$
test particles throughout the system to search for zones where earth
sized planets may reside. RL00 claim that secular resonances maintain
stability in the $\upsilon$ And system. Another group has also examined this system (Malhotra \etal 2000, Stepinski \etal 2000). They focused on the unconstrained parameters of inclination and lines of node. The simulations in Stepinski \etal show no secular resoncances in the $\upsilon$ And system. Our simulations also suggest there is no correlation between the longitudes of periastron and the stability of the system, i.e.\ the observed alignment is a coincidence.

Planet b is typical of most extra-solar planets found to date: it is of Jupiter mass and
with a small semimajor axis (0.06 AU). The new companions have
highly eccentric orbits of approximately 0.3. High eccentricities make
the stability of the system suspect. The star $\upsilon$ And is estimated to
be 2--3 billion years old; therefore, these planets should not be
transient entities. Rather than explore stability for the lifetime of
the star, we examine the overall probability that the system can be
stable on a $10^6$ year timescale, allowing a more thorough study
of parameter space. One thousand trials were run with
random initial conditions.

The studies mentioned above have only examined the $\upsilon$ And system. We decided to run a similar experiment on our Solar System to establish a fiducial point. Because we know that the Solar System is stable for $5\cdot10^9$ years, it provides a comparison system. The results of this experiment may allow stability assessments of $\upsilon$ And to be extrapolated to longer timescales.

Our methodology and results are summarized in the following sections. Section 2 is a description of the methods for generating initial
conditions and integrating the orbits.
In section 3 we present the results of the
simulations of $\upsilon$ And as well as the results of the
simulations of our own planetary system. 
Approximately 84\% of $\upsilon$
And configurations proved stable compared to 81\% for our Solar
System. Most unstable configurations ejected planet c, with stability
highly correlated to the eccentricity of planet d.  We draw some
general conclusions about these results in section 4.

\section{Numerical Methods}

The initial conditions were determined based on 
the nominal value and error for each orbital parameter as derived from
observations. A
total of 16 variables are picked for each trial. For each planet the
initial period (and hence semimajor axis), eccentricity, longitude of
periastron, and time of periastron are determined from a Gaussian
error distribution. The masses and inclinations of each planet are
degenerate. The $M_x\sin i$ value has been measured, but no estimate of the
errors in inclination can be made. Therefore the inclinations are chosen from a uniform
distribution between 0 and 5 degrees, and from this the mass is determined. This
maximum value is purely arbitrary and was chosen to encourage
stability, while still providing an adequate sampling of parameter
space. This range is in contrast to LA99 who give planet d a slight inclination and allow the inclination to dynamically evolve, and RL00 who start their simulations at 0, 30, and 60 degrees. 
The longitude of
ascending node also has no nominal value or error, hence it is 
picked from a uniform distribution between 0 and $2\pi$. The nominal
values and their
associated errors (as of 8 Sept 1999) are listed in Table 1 (Marcy,
private communication). As of 21 Jan 2000, the eccentricities of planets c and d are 0.24 and 0.31 respectively.\footnote{http://astron.berkeley.edu/$\sim$gmarcy/planetsearch/upsand/upsand.html} In the note added in proof to RL00 is a short discussion of the importance of the starting date. For these simulations the starting date is not varied, and is always chosen to be JD 2450000.00. The final piece of information is the mass of
$\upsilon$ And. This parameter is chosen from a Gaussian about $1.28\pm0.2$ $M_\odot$ (Gonzalez and Laws 2000). 

The choice of a $10^6$ year integration timescale was made to allow a
reasonable search through parameter space with limited computational
resources and is at least three orders of magnitude less than the
lifetime of the system.  The choice corresponds to 80 million orbits
for the interior planet and 280,000 orbits for the outer
planet. LA99 show that the timescale for transient terms to
die out is on the order of $10^6$ years. Therefore configurations stable for $10^6$ years should be stable for $10^8$ years, as they are in LA99. Our results do not necessarily support this optimistic theory, see \S3.1.

For comparison, simulations of the outer planets of our Solar System
were also performed. The orbital parameters of the gas giants were given
errors equal to those of the most uncertain parameters in $\upsilon$
And (typically planet d). Two sets of simulations were run, one set for
$10^6$ years, and one for $5\cdot10^6$ years.  Each set contained 32
trials. In terms of dynamical times, these simulations are of the same
order as the $\upsilon$ And simulation.

In all cases the simulations were terminated when an ejection
occurred, defined by an osculating eccentricity greater than 1.  Note
that this condition could be satisfied during a close encounter
without an actual ejection immediately ensuing.  Nevertheless, such a close
encounter bodes ill for the overall stability of the system.

The code uses a second order Mixed Variable Symplectic method as
described in Saha and Tremaine (1994; see also Wisdom and Holman
1991).   Individual timesteps are used for each planet which made the
computation much more efficient given the large difference in orbital times
between planet b and the other planets.  The stepsize for planet b was
set to 0.215 days and the ratio of the timesteps of the other plants was
1:50:200.  This corresponds to a ratio of steps per orbit of 21:22:30. The code also includes a
Hamiltonian form of general relativity in the parametrized
post-Newtonian approximation, which allowed accurate modeling of the
inner planet.  This code has been previously used in 
theoretical examinations of the stability of
our Solar System (Quinn 1998).

The advantage of symplectic integrators is that the truncation error is
equivalent to a Hamiltonian perturbation: it exactly conserves
approximate integrals of motion.  Therefore, although we are not integrating
the true system, we are integrating a Hamiltonian system that is very
similar, and which has similar stability properties.  In particular,
no secular changes in the orbits will be introduced which could
drastically affect stability.  The integrator does have two shortcomings.  First, the error in the integration increases for larger
eccentricity.  Our fixed step integrator has no mechanism to control
this error.  Second, the error in the integration can get very large
with a close encounter between two planets. This should be irrelevant
since either our termination criterion will be tripped during the close
encounter, or the errors introduced during the close encounter will
most likely make an ejection imminent, and we presume that in reality close encounters will also cause ejections.

\section{Results}
\subsection{Upsilon Andromedae}
Of the 1,000 trials, 84.0$\pm$3.4\% were stable. Three times a planet
was ejected (according to the above criterion) in less than $10^3$
years, 24 between $10^3$ and
$10^4$ years, 66 between
$10^4$ and $10^5$ years, and 67 between $10^5$ and
$10^6$ years.   Because the configurations were chosen from a
Gaussian distribution these percentages should reflect the absolute
probability that the system is stable for each timescale. This is, of course,
only true if the observational errors are also Gaussian.  Of the
160 unstable configurations, planet b was ejected 4 times, planet c
120 times, and planet d 36 times. Of the 7 simulations LA99 ran, one 
ejected planet c. Therefore they suggest the Lyapunov exponent should
be calculated based on the motion of planet c.  Our larger study
supports this hypothesis, but also reveals that the system is fully
chaotic and the motion of planet d in particular must also be
considered. Due to the huge volume of output of this set of simulations
($\sim200$ gigabytes), time resolved information was saved for only 5 trials. Only the initial and final conditions were stored for the remaining simulations. Therefore, we attempt no estimate of
the Lyapunov timescale. However, planets c and d are coupled and we expect planet d to have a similar Lyapunov time of 340 years as reported by LA99.

Because the integration time in these simulations is much shorter than the age of the star, one would like to extrapolate these numbers to the order of a gigayear. As mentioned in \S2, LA99 Figure 1 indicates that transients in the intergration damp out on the order of $10^6$ years. This would imply that configurations stable for $10^6$ years should remain stable for $10^8$ years. But the 7 simulations of LA99 are not enough to be statistically meaningful Our results show an equal number of ejections in the last two logarthmic bins. This implies a constant ejection rate per decade. We therefore encourage the reader not to draw any quantitative conclusions about the long term stability of $\upsilon$ And based on this study.. 

In general with 16 variables a principle component analysis should be
made.  However, a quick inspection indicates that for this
simulation the only parameter that determines
the short term stability is the eccentricity of planet d. Figure 1
shows how the likelihood of stability depends on
the eccentricities. All configurations in which the eccentricity
of planet d is 
less than 0.30 are stable, and all configurations in which the eccentricity is 
greater than 0.47 are unstable. Table 2 shows the likelihood of
stability as a function
of the eccentricities of both planet c and d. This table shows that in
the region between 0.27 and 0.47 the eccentricity of planet c plays a
role in stability of the system. Higher eccentricities in either c or d lead
to a higher probability for ejection.

The eccentricity of the planet d also determines the length of stability of
the system up to $10^6$ years. Although not plotted, the higher
eccentricities led to a quicker ejection. For ejections between 0 and
$10^3$ years the eccentricity of the third planet lay between 0.45 and
0.55 or +1.5 to +2.4 standard deviations from the mean. In this regime 
planet c had eccentricities between 0.3 and 0.45, also above
the mean. There is a continuous progression towards stability as the
eccentricity approaches the mean. For orbits stable up to $10^5$
years, planet d's eccentricity lay between 0.27 and 0.57 with a
mode at 0.42.

Although the eccentricity seems the critical variable, the other parameters
were also analyzed. Because of the mass/inclination degeneracy, the
effect of initial inclination needs examination. Since the
inclinations are totally unconstrained, they were chosen from a flat
distribution with maximum inclinations of 5.0 degrees.  The inclination determines the mass of each planet in our code
and hence could be the most important variable of all, but stability is almost completely independent of initial inclination. Plots of stability as a function of inclination show
only scatter about 0.84 fractional stability. Therefore the decision to
include inclinations up to 5 degrees did not impact the simulation.

Mean motion resonances appear to have little effect. The lowest order
resonances in $\upsilon$ Andromeda are near 5:1, which occur in both stable
and unstable configurations. RL00 reported that stability is highly
dependent on
the secular resonance locking of the longitudes of periastron of
planets c and d. We believe that the primary parameter that determines stability is $e_d$, therefore RL00's hypothesis is not supported by
the results presented here, or the results in Stepinski \etal We reran five stable trials and saved all the time
resolved data of the orbital parameters to examine any possible
locking mechanisms in Fourier space. In particular, we examined the
power spectra of the Poincar\'e $h$ and $k$ orbital elements. Two of
the trials do in fact show
a resonance, but these examples resided in an anti-aligned
configuration. Two of the trials showed motion resulting from the
superposition of two modes well separated in frequency, and a lower
amplitude effect
due to the inner planet. These situations are clearly not in
resonance. The fifth trial was a very chaotic system which showed
very broad band power in Fourier space. There is some indication that this
system was slightly locked in the anti-parallel configuration, however
this configuration is best described as purely chaotic. The two
resonance examples had initial $e_d$ values of 0.030 and 0.137 respectively, and
are hence much lower than the expected value.  In contrast, the cases
with two well
separated modes began
with $e_d$ values of 0.364 and 0.318.  It is also worth noting that the
simulation with $e_d = 0.318$ also had $e_c = 0.480$. The chaotic
example began with $e_c$ = 0.504 and $e_d$ = 0.275.  From these few
cases, it is apparent that 
resonance locking can occur, but 
we conclude that the current
alignment of the longitudes of periastron of planets c and d is
coincidence, and not relevant to the overall stability of the system.

\subsection{Our Solar System}
For the comparison simulations of the gas giants in our own Solar System,
the initial orbital parameters were determined from errors
equal to the largest absolute errors in the $\upsilon$ And planets. For most of the orbital parameters, planet d has the largest errors. Because of
the low eccentricities in our Solar System,
a different set of simulations using the same relative errors produced no migration or
ejections.

For the case of the $10^6$ year trials, 3 configurations
produced an ejection between $10^5$ and $10^6$ years, but the
other 29 were all stable. In the longer simulation, an ejection
occurred twice between $5\cdot10^4$ and $5\cdot10^5$ years, 11 times  between $5\cdot10^5$ and $5\cdot10^6$ years, and 19 cases were
stable. These simulations correspond to 90.6\% and 59.4\% stability
respectively. It is also useful to examine stability at the same
dynamical time. Assuming the orbital time for Jupiter and planet d is
the dynamical time, our Solar System must be integrated for
approximately 1.5 million years. This duration corresponds to 280,000
Jupiter orbits, the same as planet d in 1 million years. On this
timescale 81\% of solar system configurations are stable, which is
statistically identical to $\upsilon$ And.  Every ejection occurred
when Jupiter's initial eccentricity
was greater than 0.12 (more than +1 standard deviation), and also
followed the same inverse eccentricity-stability timescale
trend. Our solar system is stable for at least 5 billion years, yet only 81\% of the simulations were stable. This reinforces earlier results that our solar system lies on the edge of chaos (Quinn 1998, Varadi \etal 1999), and suggests that $\upsilon$ And also lies near this boundary. It is not accurate to presume that these two results imply that $\upsilon$ And is also stable for 5 billion years, but it does demonstrate that stable configurations do exist in the parameter space allowed for this system.

\section{Conclusions}

Although we show that the $\upsilon$ And planetary system formally
has an 84\% chance of being stable, the more important conclusion is
that the current observations do not provide much of a constraint on
the stability of the system.  This point is driven home by the
experiments on our own Solar System, which we show has only an
81\% chance of being stable given the same observational errors.   
The key parameter that determines 
stability is the eccentricity of planet d. Should the
eccentricity of planet d be measured to be larger than 0.47 then the
system cannot be stable under any circumstance, and the interpretation
that this is a planetary system must be rejected. Conversely, if the
eccentricity is below 0.30 the system is very likely stable for at least one
million years and longer integrations should be made to determine if
the system is stable for the lifetime of the primary. These results
are in agreement with other studies (Noyes \etal 1999, BMF, LA99, RL00).

An intriguing aspect of this study is that the best value for the
eccentricity of planet d corresponds to the edge of stability. Should
the eccentricity be any larger the system moves into an unstable
regime.  This situation is similar to what is seen in our own Solar
System, both in the sense that our planetary system may be unstable on
timescales comparable to its age (Laskar 1994), and that
relatively small changes to the planetary orbital parameters can lead
to instability on much shorter timescales (Varadi \etal 1999, Quinn 1998).
Now that we have two data points, there is a
suggestion that, in general, planetary systems reside on this precipice
of instability.  Clearly at this stage this is only a 
suggestion, but it is a possibility that could give new insights into
the nature of planet formation. This suggestion will need to be examined both as better
constraints on the orbital parameters of $\upsilon$ And become
available, and as more multiple planet systems are discovered.

\section{Acknowledgements}
This research would not have been possible without the assistance of
Geoff Marcy, Joachim Stadel, George Lake, Derek Richardson, Frank van
den Bosch, and Chance Reschke.  We would also like to thank Prasenjit
Saha for providing his code from which the integration code used here
was derived.  This work was partially supported from NASA grant NAG5-3462
and from NSF grant 9979891.  The simulations were performed on
workstations donated by Intel Corporation.

\newpage

Fig. 1 - The dependence of stability on eccentricity. This
histogram shows the fraction of stable orbits binned by the
initial eccentricity of each planet.  Binning by Planet b's
eccentricity is represented by x's, c's by open
triangles, and d's by squares joined by a line. Bin sizes vary by planet due to
different ranges of possible values, but all are normalized based on
the mean and standard deviation. 
Note that the points are uncorrelated, i.e.\ if the eccentricity of planet d is 0, the other eccentricities could be any value.

\newpage
\begin{center}Table 1 - Observational Values and Errors for Upsilon Andromeda
\end{center}
\begin{tabular}{c c c c c c}
\hline \hline
Planet & Eccentricity & Period(days) & Long of Peri.(degrees) & Time of Peri.(JD) & Msin i($M_J$) \\
\hline 
b & 0.025$\pm$0.015 & 4.6171$\pm$0.0003 & 83.0$\pm$243.0 & 2450001.0$\pm$3.1 & 0.71 \\
c & 0.29$\pm$0.11 & 241.02$\pm$1.1 & 243.6$\pm$33.0 & 2450159.8$\pm$20.8 & 2.11 \\
d & 0.29$\pm$0.11 & 1306.59$\pm$30.0 & 247.7$\pm$17.0 & 2451302.6$\pm$40.6 & 4.61 \\
\hline
\end{tabular}

\end{document}